\documentclass[aps,prb,twocolumn]{revtex4}

\usepackage{graphicx}
\usepackage{tabularx}
\usepackage{amsfonts}
\usepackage{amssymb}
\usepackage{amsbsy}
\begin{document}

\title{Disorder mediated splitting of the cyclotron resonance in two-dimensional electron systems}

\author{E. A. Henriksen}
\affiliation{Department of Physics, Columbia University, New York, NY 10027}
\author{S. Syed}
 \altaffiliation[Present address: ]{Department of Physics, University of Illinois, Urbana-Champaign, Urbana, IL 61801}
\affiliation{Department of Physics, Columbia University, New York, NY 10027}
\author{Y.-J. Wang}
\affiliation{National High Magnetic Field Lab, Florida State University, Tallahassee, FL 32310}
\author{H. L. Stormer}
\affiliation{Department of Physics, Columbia University, New York, NY 10027}
\affiliation{Bell Laboratories, Lucent Technologies, Murray Hill, NJ 07974}
\author{L. N. Pfeiffer}
\affiliation{Bell Laboratories, Lucent Technologies, Murray Hill, NJ 07974}
\author{K. W. West}
\affiliation{Bell Laboratories, Lucent Technologies, Murray Hill, NJ 07974}
\begin{abstract}
We perform a direct study of the magnitude of the anomalous splitting in the cyclotron resonance (CR) of a two-dimensional electron system (2DES) as a function of sample disorder.  In a series of AlGaAs/GaAs quantum wells, identical except for a range of carbon doping in the well, we find the CR splitting to vanish at high sample mobilities but to increase dramatically with increasing impurity density and electron scattering rates.  This observation lends strong support to the conjecture that the non-zero wavevector, roton-like minimum in the dispersion of 2D magnetoplasmons comes into resonance with the CR, with the two modes being coupled via disorder.
\end{abstract}
\maketitle

Splittings of the far-infrared cyclotron resonance (FIR CR) line frequently occur in two-dimensional electron systems (2DESs) due to an anti-crossing of this well-established, magnetic field (B) dependent resonance with other modes of the system.  In most cases, optical phonons \cite{Wang97} or transitions between electronic sub-bands \cite{Schlesinger83} of the 2DES are the second excitation of the solid that comes into resonance with the CR energy to create the characteristic splitting pattern.  However, a CR splitting resembling such an anti-crossing, first observed by Schlesinger $\it{et~al}$ \cite{Schlesinger84} in AlGaAs/GaAs heterostructures, has withstood such an interpretation since no obvious second excitation of the semiconductors seems to fit the experimental data.  Instead, the observation that the splitting energy is proportional to the Coulomb interaction energy E$_{coul} = e^2 \sqrt{n_{2D}} / \epsilon$, where $n_{2D}$ is the 2D electron density and $\epsilon$ is the static dielectric constant, suggests that the CR is coupling to excitations of the interacting 2DES.  The recent observation of a similar splitting in the 2DES of the AlGaN/GaN material system \cite{Syed03} underscores the ubiquity of this phenomenon and the importance of resolving the nature of its origin.

In this communication, we establish a strict dependence of the magnitude of this anomalous splitting of the CR line on sample disorder.  In a sequence of AlGaAs/GaAs 2DESs, which are identical except for a different level of impurities in each sample, we observe the size of the splitting to strongly increase with impurity concentration and electron scattering rates.  According to Kohn's theorem \cite{Kohn61}, in a system with translational symmetry the CR lineshape is unaffected by electron-electron interactions.  However, the presence of disorder breaks this symmetry and can mix the CR with non-zero wavevector modes.  Our results provide quantitative support for such a condition in our samples, and we conclude that an excitation of the interacting 2DES $\it{away}$ from $k = 0$ and degenerate with the CR is at the origin of this puzzling anti-crossing.  Sample disorder provides the scattering wavevectors necessary to couple this mode to the CR, such that more disorder leads to stronger coupling.

Previous investigations raised the possibility of such a coupling as the origin of the splitting of the CR line.  Measurements by Syed $\it{et~al}$ in a variety of AlGaAs/GaAs 2D samples indicated that the splitting size was correlated with the sample mobility \cite{Syed04.2}.  However, the widely varying sample parameters and sources of electron scattering in these samples prevented the observation of an actual dependence.  Richter $\it{et~al}$ established that only scattering from acceptor impurities, and not donor impurities, is responsible for the splitting in AlGaAs/GaAs heterostructures \cite{Richter89}.  They realized that samples with greater disorder showed stronger CR anomalies.  However, over the narrow range of mobilities available to them, and with the widely varying range of electron densities in their samples, no study of any dependence of the splitting size on disorder was made.  Zhao $\it{et~al}$ studied CR spectra in AlGaAs/GaAs samples having surface metal gratings with varying periods.  While they observed splittings in the grating-induced 2D plasmon dispersion due to crossing with the second harmonic of the CR, a splitting of the CR itself was only seen in one single low mobility sample, and no study of its dependence on any parameters was made \cite{Zhao95}.  The impact of sample mobility on $\it{multiple}$ CR line splittings observed in Si inversion layers was studied by Cheng and McCombe \cite{Cheng91}.  They noted that lower mobility Si inversion layers exhibited more peaks, with greater peak-to-peak separation and generally clearer structure than in high mobility samples.  The authors proposed a number of competing effects that can combine to create the observed complicated behavior.  Yet the complexity of the Si band structure and the multiple sources of electron scattering precludes a clear identification of the origin of the splitting in this material system.

\begin{figure}
\includegraphics[width=0.47\textwidth]{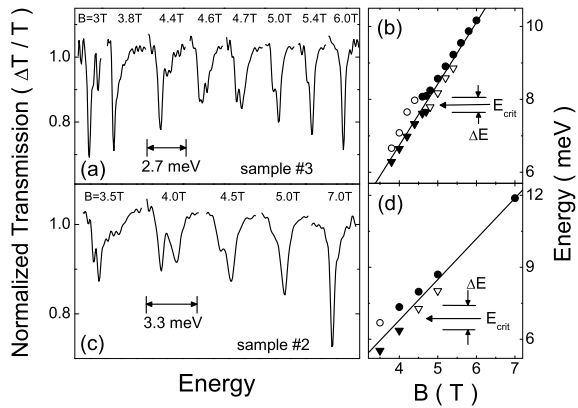}
\caption{\label{Fig.1} (a) and (c): Normalized CR transmission ($\Delta T / T$) for a selection of B fields. Traces are separated to avoid overlap and enhance clarity.  Panels (b) and (d) show energy position of $\Delta T / T$ minima vs. B field; the solid line is the CR expected from the high and low field data.  Circles (triangles) show higher (lower) energy minima, and closed (open) symbols indicate the stronger (weaker) minima.  Data in (a) and (b) are from sample \#3 while (c) and (d) are from sample \#2.  System noise increases at lower energies.}
\end{figure}

In order to elucidate the role of disorder in the splitting of CR in 2DESs, we have performed systematic experiments on the effect of electron scattering from acceptor impurities on the CR lineshape.  This study is performed in AlGaAs/GaAs, which offers the best control over the scattering properties of its carriers.  We introduced carbon acceptors, C, during molecular beam epitaxy (MBE) growth into the GaAs channel of modulation-doped AlGaAs/GaAs quantum wells.  Except for varying the C impurity density, $N_{imp}$, all samples are otherwise identical, having a 300 \AA {} wide GaAs well embedded in Al$_{0.33}$Ga$_{0.67}$As with single-sided doping at a setback of 800 \AA.  Values for $N_{imp}$ were calibrated by doping bulk GaAs with the same impurity flux, and determining the impurity concentration via Hall measurements.  The 2D impurity density, $n_{imp}$, in the GaAs channel of our samples is determined as $n_{imp} = N_{imp} \times 300$ \AA {} and ranges from $2\times10^9$ cm$^{-2}$ to $3.2\times10^{10}$ cm$^{-2}$.  The resulting samples have mobilities, $\mu$,  from $1.5\times10^6$ cm$^2$/Vs to $4\times10^4$ cm$^2$/Vs respectively, but with only a narrow range of 2D electron densities, $n_{2D}$, from $1.8 \times 10^{11}$ to $2.4 \times 10^{11}$ cm$^{-2}$.  The mobility of identical square wells in the absence of C doping is larger than $5\times10^6$ cm$^2$/Vs, demonstrating the very low levels of background impurities in these samples.  Electrical contacts were made by annealing indium beads at the edges of the $4\times4$ mm$^2$ samples.

\begin{table}[b]
\caption{\label{Table 1} Parameters of all AlGaAs/GaAs C-doped quantum wells used in our experiments.  The mobility, $\mu$, is in units of 10$^4$ cm$^2$/Vs; $n_{2D}$ and $n_{imp}$ are in 10$^{10}$ cm$^{-2}$; $E_{crit}$ and $\Delta E$ are in meV; $\nu$ is the Landau level filling factor at $B_{crit}$.  Parantheses indicate broadenings (see text).} 
\begin{tabular*}{0.45\textwidth}{@{\extracolsep{\fill}}|ccccccc|}
\hline
Sample & $\mu$ & $n_{2D}$ & $n_{imp}$ & $E_{crit}$ & $\Delta E$ & $\nu$ \\
\hline
1 & 4.1 & 20 & 3.2 & 6.1 & 1.2 & 2.2 \\
2 & 9.4 & 18 & 2.0 & 7.0 & 0.98 & 1.7 \\
3 & 26.6 & 24 & 0.8 & 7.9 & 0.42 & 2.4 \\
4 & 51.0 & 23 & 0.4 & 7.6 & 0.39 & 2.7 \\
5 & 74.0 & 21 & 0.3 & 7.4 & (0.25) & 1.9 \\
6 & 155 & 20 & 0.2 & 6.5 & (0.07) & 1.9 \\
\hline
\end{tabular*}
\end{table}

\begin{figure*}[t]
\includegraphics[width=\textwidth]{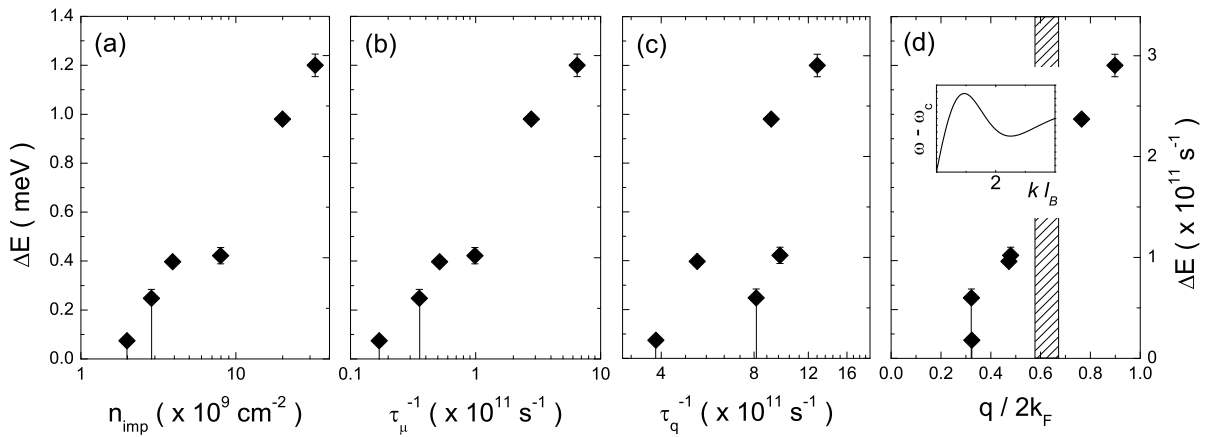}
\caption{\label{Fig.2} $\Delta E$ as a function of (a) 2D impurity density, (b) transport scattering rate, (c) quantum scattering rate, and (d) $q / 2 k_F$.  Samples \#5 and \#6 show a broadening rather than a splitting and so have error bars extending to $\Delta E =  0$.  The hashed area in (d) mark the range of the $k \approx 2 / l_B$ magnetoroton minimum at $B_{crit}$, for all samples.  The inset to (d) shows the 2D magnetoplasmon dispersion at filling factor $\nu = 2$.}
\end{figure*}

CR experiments were conducted at 4.2 K in magnetic fields up to 15T, using a Bruker IFS 113v Fourier transform spectrometer in conjunction with light pipe optics and a composite Si bolometer.  The samples were backside-wedged to $\approx 10^{\circ}$ to suppress Fabry-Perot interference.  Standard four-terminal lock-in techniques were used for simultaneous $\it{in ~ situ}$ transport measurements.  All data were taken after the samples were illuminated with a red LED.  Additional measurements of sample magnetoresistance were made in separate cooldowns.  Table 1 displays the transport parameters and results extracted from the CR measurements. 

Fig. 1 shows representative sets of transmission measurements ($\Delta T / T$) on two samples for selected values of the applied magnetic field, $B$.  In all samples, spectra were taken with 0.25 meV resolution and normalized to scans at $B = 0$ T.  Clear splittings of the CR line were observed in four of our samples at magnetic fields around 4-5T, while the two highest mobility samples showed only a characteristic broadening.  All traces were fit to Lorentzians, with two curves used in the range of the  splittings and broadenings.  In all cases the fits provided good matches to the data.  We characterize the splittings by $B_{crit}$, the magnetic field where the two fitting Lorentzians have equal intensities; by $\Delta E$, the distance between these Lorentzians; and by $E_{crit}$, the ``center-of-mass'' energy of the two resonances at $B_{crit}$.  We emphasize that the splitting can be as large as $\sim$ 20\% of the cyclotron energy, comparable to the AlGaN/GaN results \cite{Syed03}.  

In Fig. 1 (a) and (c), a splitting is clearly visible in the CR lines.  Outside the range of the splitting, the CR field dependence yields the standard electron effective mass of $m^* = 0.068 m_e$.  For  sample \#3, with increasing field above $B = 3.8$ T, the CR is shifted to $\omega \textless \omega_c$, where $\omega_c = e B / m^*$ is the cyclotron frequency, with a second, weaker and broader, resonance in the transmission appearing at  $\omega \textgreater \omega_c$.  As $B$ is increased further, the intensity of the higher energy resonance grows at the expense of the lower until the two are of roughly equal intensities at $B = 4.6$ T, where we find $E_{crit} = 7.85$ meV and $\Delta E = 0.42$ meV.  With further increase in $B$, the lower resonance is supplanted by the higher one, which is shifted back to $\omega_c$.  This motion can be seen clearly in Fig. 1(b), which depicts the position of the resonances versus $B$ field.  At $B_{crit}$ the two resonances have approximately the same intensity.  On either side of $B_{crit}$ the intensities become increasingly unequal, with strong (weak) relative intensity indicated by closed (open) symbols. Similar behavior is displayed by the lower mobility sample \#2 in Fig. 1 (c) and (d), except that relative to sample \#3 the resonances are wider, and the splitting is larger by more than a factor of two, at $\Delta E = 0.98$ meV.  Across all samples, the mobility has a pronounced effect on the widths of the resonances both at the splitting and at fields $B < B_{crit}$.  Lower sample mobilities are correlated with wider resonances, with their FWHM in reasonable agreement with values calculated from the self-consistent Born approximation \cite{Ando75}, $\Delta \omega^2 = ( 2 / \pi ) \omega_c / \tau_{\mu}$, where $\tau_{\mu}$ is the transport scattering time, covering an order of magnitude change in width (1.7 to 0.2 meV) over the entire range of sample mobilities.  However at high fields, close to filling factor $\nu = 1$, the widths are generally narrow and vary little (0.3 to 0.2 meV) over the mobility range \cite{Richter89, Zaremba91}.

We determined the transport parameters of our samples via standard van der Pauw and Hall measurements, and their impact on $\Delta E$ is shown in Fig. 2.  Fig. 2 (a) shows $\Delta E$ as a function of the C impurity density, $n_{imp}$ and in Fig. 2 (b) we plot $\Delta E$ against the transport scattering rate $\tau_{\mu}^{-1}$ derived from the mobility, $\mu = e \tau_{\mu} / m^*$.  We also performed Shubnikov-de Haas (SdH) measurements on our samples to obtain the dependence of the splitting on the quantum scattering rate, $\tau_q^{-1}$.  The average lifetime of a momentum eigenstate, $\tau_q$, is determined by constructing a Dingle plot from the SdH data \cite{Coleridge91}.  Fig. 2 (c) shows $\Delta E$ as a function of $\tau_q^{-1}$.

The panels of Fig. 2 represent the central findings of our experiments. All data show a clear and rapid increase of the CR splitting, $\Delta E$, with sample disorder. The dependence of $\Delta E$ on the various parameters in Fig. 2 is roughly linear on these semilog plots, seen most clearly in the $n_{imp}$ and $\tau_{\mu}^{-1}$ dependences.  The $\Delta E$ vs. $\tau_q^{-1}$ data show more scatter, which is probably due to the more complex analysis required for SdH than for Hall and van der Pauw measurements, and may also be due to the fact that separate cooldowns were required to collect the SdH data, while Hall and van der Pauw data were collected together with the CR data.

Further exploiting the content of our data sets we can derive the average scattering wavevector $q$ from a combination of $\tau_{\mu}$ and $\tau_q$.   Modeling low-temperature electron scattering about the 2D Fermi surface at $k_F$ as a random walk, with each step taking time $\tau_q$ to scatter through an angle $\phi$ and reaching 180$^{\circ}$ in time $\tau_{\mu}$, one can derive \cite{Syed04} the average scattering wavevector $q$ to be

\begin{displaymath}
q = 2 k_F sin(\phi/2) = 2 k_F sin\left( \frac{\pi}{2}\sqrt{\frac{\tau_q}{\tau_{\mu}}}\right).
\end{displaymath}

This $q$ represents the ``decay length'' of a distribution of scattering wavevectors extending into $k$-space from $k=0$. Fig 2 (d) shows $\Delta E$ as a function of $q / 2 k_F$.  Clearly, larger $q$ values, representing a stronger scattering potential, are correlated with larger splittings.

The data in Fig. 1 (b) and (d) are consistent with an anti-crossing between the CR mode and a second, energetically degenerate mode, whereby the increased splitting is a result of increased coupling strengths.  In the context of our experiments, this coupling is due to the disorder potential, where the primary effect of disorder is to break the translational invariance leading to a smearing of the electron momentum.  Such smearing causes the mixing of modes at different points in momentum space, and is particularly effective for modes at extrema in the dispersion of the 2DES.  Therefore we infer the existence of a mode, degenerate with the CR but removed in $k$-space, which is increasingly coupled to the CR by an increasing amount of sample disorder.

Several semiconductor modes that are candidates to resonate with the CR, including optical or acoustic phonons, impurity or subband transition, etc., can either be ruled out on an energy basis (e.g. $E_{crit}$ is too low to excite optical phonons) or by direct examination.  For instance, in sample \#2 we have verified that tilting the sample by $10^{\circ}$ causes an additional level anti-crossing in the CR at B = 12 T that, although similar in appearance, is due to mixing of the CR with subband transitions of the quantum well \cite{Schlesinger83}.  However the tilt leaves the CR splitting under investigation unchanged.  A remaining candidate for the interacting mode is the magnetoroton, a roton-like minimum in the dispersion of 2D magnetoplasmons (see inset to Fig 2 (d) )\cite{Schlesinger84, Kallin85}.  This mode satisfies two of the requirements for the second mode: it is an extremum of the magnetoplasmon dispersion and has a non-zero wavevector $k \approx 2 / l_B$, where $l_B = \sqrt{\hbar c / e B}$ is the magnetic length.  The range of wavevectors of such magnetorotons in our set of samples is indicated by the hashed area in Fig. 2 (d), showing that the distribution of characteristic scattering wavevectors overlaps strongly with the magnetoroton in momentum space.  On the other hand, while the magnetoplasmon dispersion is anchored to the CR energy for $k = 0$ \cite{Kohn61}, the magnetoroton energy lies above the CR\cite{Kallin85, MacDonald85, Oji86}, although higher order magnetoplasma-magnetoplasma scattering\cite{Cheng94} or fluctuations of the center of mass of the 2DES\cite{Hu88} may lower it into degeneracy with the CR.  In addition, the data of Fig. 1 imply that the second mode energy lies above the CR energy at low fields, and $\it{decreases}$ relative to $\hbar \omega_c$ with increasing field.  In contrast, calculations of the magnetoroton energy typically find it to increase $\it{away}$ from the CR energy with increasing field\cite{Kallin85, MacDonald85}.

As demonstrated by our results, sample disorder clearly plays a central role in the observation of this CR splitting, with the largest splittings linked to the strongest scattering by impurities.  Within the magnetoplasmon model, theory usually treats an ideal magnetoplasmon dispersion coupled to the FIR CR via a weak disorder potential composed of typical electron scattering sources: remote ionized donors, neutral background impurities, or interface roughness \cite{Kallin85}.  However in our lowest mobility samples, the impurity broadening is a sizeable fraction of the Coulomb interaction energy of the 2DES at fields near the splitting, and it becomes questionable whether weak coupling is still assured.  Further, experimental evidence suggests that only disorder composed of negatively charged impurities creates the observed splittings \cite{Sigg88, Richter89}.  These two issues have been addressed in some detail.  In the low-disorder limit Kallin and Halperin find a weak, broad peak above the CR due to coupling of the $k = 0$ CR with the non-zero wavevector magnetoroton via short-range impurity scattering.  However, no splitting is found because the two modes are never degenerate.  Antoniou and MacDonald\cite{Antoniou92} have studied the interaction of magnetoplasmons and CR for an extended range of disorder strength, although only for short-range scatterers and at $\nu = 1$ and $0.5$.  In their model weak disorder also induces a very broad second bump above the CR, which becomes dominant in the strong disorder limit.  However, the paper does not provide evidence for the appearance of an anti-level crossing as a function of varying magnetic field.  Zaremba has shown\cite{Zaremba91} that acceptor and donor impurities can have strikingly different effects on the CR of 2D electrons, as seen experimentally.  However, neither creates a splitting in this model.  From these discussions we conclude that the origin of this anomalous splitting in the CR remains unresolved.

In conclusion, we have performed a systematic study of the disorder dependence of the anomalous splitting of the CR line in 2D systems.  In a sequence of AlGaAs/GaAs quantum wells in which acceptor impurities were intentionally incorporated into the 2D channel, we find a large increase of the CR splitting size with increasing impurity concentration and associated electron scattering rates.  The exact origin of this splitting remains unclear.  However our results provide strong support for a model of the splitting as an anti-crossing between the CR and a second, energetically degenerate mode at non-zero wavevector, coupled via electron scattering and the ensuing lack of momentum conservation.

\begin{acknowledgments}
The authors wish to thank C. Kallin, D. Smirnov, Y. Ahmadian, A. Mitra, C. Hirjibeheddin, and A. Pinczuk for valuable discussions. Funding under ONR Project no. N00014-04-1-0028 is acknowledged. A portion of this work was performed at the National High Magnetic Field Laboratory, which is supported by NSF Cooperative Agreement No. DMR-0084173 and the State of Florida.  Financial support from the W. M. Keck Foundation is gratefully acknowledged.
\end{acknowledgments}


\begin{thebibliography}{15}
\expandafter\ifx\csname
natexlab\endcsname\relax\def\natexlab#1{#1}\fi
\expandafter\ifx\csname bibnamefont\endcsname\relax
  \def\bibnamefont#1{#1}\fi
\expandafter\ifx\csname bibfnamefont\endcsname\relax
  \def\bibfnamefont#1{#1}\fi
\expandafter\ifx\csname citenamefont\endcsname\relax
  \def\citenamefont#1{#1}\fi
\expandafter\ifx\csname url\endcsname\relax
  \def\url#1{\texttt{#1}}\fi
\expandafter\ifx\csname
urlprefix\endcsname\relax\def\urlprefix{URL }\fi
\providecommand{\bibinfo}[2]{#2}
\providecommand{\eprint}[2][]{\url{#2}}

\bibitem{Wang97}
\bibinfo{author}{\bibfnamefont{Y.~J.}~\bibnamefont{Wang}},
\bibinfo{author}{\bibfnamefont{H.~A.}~\bibnamefont{Nickel}},
\bibinfo{author}{\bibfnamefont{B.~D.}~\bibnamefont{McCombe}},
\bibinfo{author}{\bibfnamefont{F.~M.}~\bibnamefont{Peeters}},
\bibinfo{author}{\bibfnamefont{J.~M.}~\bibnamefont{Shi}},
\bibinfo{author}{\bibfnamefont{G.~Q.}~\bibnamefont{Hai}},
\bibinfo{author}{\bibfnamefont{X.-G.}~\bibnamefont{Wu}},
\bibinfo{author}{\bibfnamefont{T.~J.}~\bibnamefont{Eustis}},
\bibinfo{author}{\bibfnamefont{W.}~\bibnamefont{Schaff}},
  \bibinfo{journal}{Phys. Rev. Lett.} \textbf{\bibinfo{volume}{79}},
  \bibinfo{pages}{3226} (\bibinfo{year}{1997}).
  
\bibitem{Schlesinger83}
\bibinfo{author}{\bibfnamefont{Z.}~\bibnamefont{Schlesinger}},
\bibinfo{author}{\bibfnamefont{J.~C.~M.}~\bibnamefont{Hwang}},
\bibinfo{author}{\bibfnamefont{S.~J.}~\bibnamefont{Allen}},
  \bibinfo{journal}{Phys. Rev. Lett.} \textbf{\bibinfo{volume}{50}},
  \bibinfo{pages}{2098} (\bibinfo{year}{1983}).

\bibitem{Schlesinger84}
\bibinfo{author}{\bibfnamefont{Z.}~\bibnamefont{Schlesinger}},
\bibinfo{author}{\bibfnamefont{S.~J.}~\bibnamefont{Allen}},
\bibinfo{author}{\bibfnamefont{J.~C.~M.}~\bibnamefont{Hwang}},
\bibinfo{author}{\bibfnamefont{P.~M.}~\bibnamefont{Platzman}},
\bibinfo{author}{\bibfnamefont{N.}~\bibnamefont{Tzoar}},
  \bibinfo{journal}{Phys. Rev. B} \textbf{\bibinfo{volume}{30}},
  \bibinfo{pages}{R435} (\bibinfo{year}{1984}).
  
  \bibitem{Syed03}
\bibinfo{author}{\bibfnamefont{S.}~\bibnamefont{Syed}},
\bibinfo{author}{\bibfnamefont{M.~J.}~\bibnamefont{Manfra}},
\bibinfo{author}{\bibfnamefont{Y.~J.}~\bibnamefont{Wang}},
\bibinfo{author}{\bibfnamefont{H.~L.}~\bibnamefont{Stormer}},
\bibinfo{author}{\bibfnamefont{R.~J.}~\bibnamefont{Molnar}},
  \bibinfo{journal}{Phys. Rev. B} \textbf{\bibinfo{volume}{67}},
  \bibinfo{pages}{241304(R)} (\bibinfo{year}{2003}).

\bibitem{Kohn61}
\bibinfo{author}{\bibfnamefont{W.}~\bibnamefont{Kohn}},
  \bibinfo{journal}{Phys. Rev.} \textbf{\bibinfo{volume}{123}},
  \bibinfo{pages}{1242} (\bibinfo{year}{1961}).

\bibitem{Syed04.2}
\bibinfo{author}{\bibfnamefont{S.}~\bibnamefont{Syed}},
\bibinfo{author}{\bibfnamefont{Y.~J.}~\bibnamefont{Wang}},
\bibinfo{author}{\bibfnamefont{H.~L.}~\bibnamefont{Stormer}},
\bibinfo{author}{\bibfnamefont{M.~J.}~\bibnamefont{Manfra}},
\bibinfo{author}{\bibfnamefont{L.~N.}~\bibnamefont{Pfeiffer}},
\bibinfo{author}{\bibfnamefont{K.~W.}~\bibnamefont{West}},
\bibinfo{author}{\bibfnamefont{R.}~\bibnamefont{Molnar}},
  \bibinfo{journal}{Intl. J. Mod. Phys. B} \textbf{\bibinfo{volume}{18}},
  \bibinfo{pages}{3761} (\bibinfo{year}{2004}).

\bibitem{Richter89}
\bibinfo{author}{\bibfnamefont{J.}~\bibnamefont{Richter}},
\bibinfo{author}{\bibfnamefont{H.}~\bibnamefont{Sigg}},
\bibinfo{author}{\bibfnamefont{K.~v.}~\bibnamefont{Klitzing}},
\bibinfo{author}{\bibfnamefont{K.}~\bibnamefont{Ploog}},
  \bibinfo{journal}{Phys. Rev. B} \textbf{\bibinfo{volume}{39}},
  \bibinfo{pages}{R6268} (\bibinfo{year}{1989}).

\bibitem{Zhao95}
\bibinfo{author}{\bibfnamefont{Y.}~\bibnamefont{Zhao}},
\bibinfo{author}{\bibfnamefont{D.~C.}~\bibnamefont{Tsui}},
\bibinfo{author}{\bibfnamefont{M.~B.}~\bibnamefont{Santos}},
\bibinfo{author}{\bibfnamefont{M.}~\bibnamefont{Shayegan}},
\bibinfo{author}{\bibfnamefont{R.~A.}~\bibnamefont{Ghanbari}},
\bibinfo{author}{\bibfnamefont{D.~A.}~\bibnamefont{Antoniadis}},
\bibinfo{author}{\bibfnamefont{H.~I.}~\bibnamefont{Smith}},
  \bibinfo{journal}{Phys. Rev. B} \textbf{\bibinfo{volume}{51}},
  \bibinfo{pages}{13174} (\bibinfo{year}{1995}).

\bibitem{Cheng91}
\bibinfo{author}{\bibfnamefont{J.-P.}~\bibnamefont{Cheng}},
\bibinfo{author}{\bibfnamefont{B.~D.}~\bibnamefont{McCombe}},
  \bibinfo{journal}{Phys. Rev. B} \textbf{\bibinfo{volume}{44}},
  \bibinfo{pages}{3070} (\bibinfo{year}{1991}).

\bibitem{Ando75}
\bibinfo{author}{\bibfnamefont{T.}~\bibnamefont{Ando}},
  \bibinfo{journal}{J. Phys. Soc. Jpn.} \textbf{\bibinfo{volume}{38}},
  \bibinfo{pages}{989} (\bibinfo{year}{1975}).

\bibitem{Coleridge91}
\bibinfo{author}{\bibfnamefont{P.~T.}~\bibnamefont{Coleridge}},
  \bibinfo{journal}{Phys. Rev. B} \textbf{\bibinfo{volume}{44}},
  \bibinfo{pages}{3793} (\bibinfo{year}{1991}).

\bibitem{Syed04}
\bibinfo{author}{\bibfnamefont{S.}~\bibnamefont{Syed}},
\bibinfo{author}{\bibfnamefont{M.~J.}~\bibnamefont{Manfra}},
\bibinfo{author}{\bibfnamefont{Y.~J.}~\bibnamefont{Wang}},
\bibinfo{author}{\bibfnamefont{R.~J.}~\bibnamefont{Molnar}},
\bibinfo{author}{\bibfnamefont{H.~L.}~\bibnamefont{Stormer}},
  \bibinfo{journal}{Appl. Phys. Lett.} \textbf{\bibinfo{volume}{84}},
  \bibinfo{pages}{1507} (\bibinfo{year}{2004}).
  
\bibitem{Kallin85}
\bibinfo{author}{\bibfnamefont{C.}~\bibnamefont{Kallin}},
\bibinfo{author}{\bibfnamefont{B.~I.}~\bibnamefont{Halperin}},
  \bibinfo{journal}{Phys. Rev. B} \textbf{\bibinfo{volume}{31}},
  \bibinfo{pages}{3635} (\bibinfo{year}{1985}).
  
\bibitem{MacDonald85}
\bibinfo{author}{\bibfnamefont{A.~H.}~\bibnamefont{MacDonald}},
  \bibinfo{journal}{J. Phys. C} \textbf{\bibinfo{volume}{18}},
  \bibinfo{pages}{1003} (\bibinfo{year}{1985}).
  
\bibitem{Oji86}
\bibinfo{author}{\bibfnamefont{H.~C.~A.}~\bibnamefont{Oji}},
\bibinfo{author}{\bibfnamefont{A.~H.}~\bibnamefont{MacDonald}},
  \bibinfo{journal}{Phys. Rev. B} \textbf{\bibinfo{volume}{33}},
  \bibinfo{pages}{3810} (\bibinfo{year}{1986}). 

\bibitem{Cheng94}
\bibinfo{author}{\bibfnamefont{S.-C.}~\bibnamefont{Cheng}},
  \bibinfo{journal}{Phys. Rev. B} \textbf{\bibinfo{volume}{49}},
  \bibinfo{pages}{4703} (\bibinfo{year}{1994}).

\bibitem{Hu88}
\bibinfo{author}{\bibfnamefont{G.~Y.}~\bibnamefont{Hu}},
\bibinfo{author}{\bibfnamefont{R.~F.}~\bibnamefont{O'Connell}},
  \bibinfo{journal}{Phys. Rev. B} \textbf{\bibinfo{volume}{37}},
  \bibinfo{pages}{R10391} (\bibinfo{year}{1988}).

\bibitem{Sigg88}
\bibinfo{author}{\bibfnamefont{H.}~\bibnamefont{Sigg}},
\bibinfo{author}{\bibfnamefont{D.}~\bibnamefont{Weiss}},
\bibinfo{author}{\bibfnamefont{K.~von}~\bibnamefont{Klitzing}},
  \bibinfo{journal}{Surf. Sci.} \textbf{\bibinfo{volume}{196}},
  \bibinfo{pages}{293} (\bibinfo{year}{1988}).
  
\bibitem{Antoniou92}
\bibinfo{author}{\bibfnamefont{D.}~\bibnamefont{Antoniou}},
\bibinfo{author}{\bibfnamefont{A.~H.}~\bibnamefont{MacDonald}},
  \bibinfo{journal}{Phys. Rev. B} \textbf{\bibinfo{volume}{46}},
  \bibinfo{pages}{15225} (\bibinfo{year}{1992}).
 
\bibitem{Zaremba91}
\bibinfo{author}{\bibfnamefont{E.}~\bibnamefont{Zaremba}},
  \bibinfo{journal}{Phys. Rev. B} \textbf{\bibinfo{volume}{44}},
  \bibinfo{pages}{R1379} (\bibinfo{year}{1991}).
 
 \end{thebibliography}
\end{document}